\documentclass[a4paper,twoside,11pt]{article}
\usepackage[cp1251]{inputenc}
\usepackage[english,russian]{babel}
\usepackage{fancyhdr}
\usepackage{newprog1e}
\usepackage{amsfonts,amsmath,amssymb,amsthm}
\usepackage{graphicx}

\numberwithin{equation}{section}

\journalnumber{}
\curyear{2018}
\authorlist{КРЮКОВ, ДЕМИЧЕВ}
\titlehead{ДЕЦЕНТРАЛИЗОВАННЫЕ ХРАНИЛИЩА ДАННЫХ}
\headerdef

\udk{004.75}
\rubrika{}
\dateinput{}

\rusabstr{В работе представлен сравнительный обзор децентрализованных хранилищ данных разных типов. Показано, что хотя они обладают рядом общих свойств, характерных для всех одноранговых (P2P) информационно-вычислительных сетей, решаемые задачи и, соответственно, используемые технологии построения хранилищ разных типов существенно различаются.
}

\author{
{\bfseries  А.~П.~Крюков, А.~П.~Демичев}
\\ {\itshape Федеральное государственное бюджетное образовательное учреждение высшего образования «Московский государственный университет имени М.В.Ломоносова», Научно-исследовательский институт ядерной физики имени Д.В.Скобельцына}
\\ {\slshape 119991} {\itshape ГСП-1, Москва, Ленинские горы, } {\slshape 1, стр.2}
\\ {\itshape E-mail: kryukov@theory.sinp.msu.ru, demichev@sinp.msu.ru}}
\title{ДЕЦЕНТРАЛИЗОВАННЫЕ ХРАНИЛИЩА ДАННЫХ: ТЕХНОЛОГИИ ПОСТРОЕНИЯ\thanks{Работа, представленная в разделах~\ref{sec:DHP} и \ref{sec:PKR-PUB}, проводилась при финансовой поддержке РНФ (грант № 18-41-06003). Работа, представленная в разделах~\ref{sec:DHPD} и \ref{sec:PKR-PRIV}, проводилась при финансовой поддержке РНФ (грант № 18-11-00075).}}

\date{}

\begin{document}

\maketitle
\setcounter{page}{1}
\He{ВВЕДЕНИЕ}

В настоящее время во всем мире бурно развивается исследование методов построения, разработка и использование информационно-вычислительных одноранговых (Peer-to-Peer; P2P) сетей \cite{1}. Основные преимущества P2P-сетей состоят в том, что они не требуют специального администрирования, адаптивны, участники могут свободно присоединяться и покидать сеть. Очень важно, что они  могут объединить и использовать огромные вычислительные ресурсы и ресурсы хранения посредством интернета, распределены, децентрализованы и поэтому потенциально отказоустойчивы и осуществляют балансировку нагрузки. Чаще всего такие сети используются для передачи аудио- и видеоконтента, а также для организации децентрализованных хранилищ (ДЦХ) данных. Кроме этого, такие сети используются для параллельных вычислений, распределенного кэширования ресурсов, создание систем, устойчивых к атакам типа ``отказ в обслуживании'', распространения программных модулей и ряда других задач. Основные преимущества P2P-сетей состоят в том, что они \cite{2}:
\begin{itemize}
\item не требуют специального администрирования (zero administration approach);
\item обладают возможностями самоорганизации и адаптивности; пиры (то есть, участники сети) могут свободно присоединяться и покидать сеть, P2P-системы обрабатывают эти события автоматически;
\item могут объединить и использовать огромные вычислительные ресурсы и ресурсы хранения посредством интернета;
\item распределены и децентрализованы; поэтому они потенциально отказоустойчивы и обладают свойством самобалансировки нагрузки;
\item некоторые P2P-сети имеют встроенные механизмы обеспечения анонимности пользователей (мы не рассматриваем этот вопрос в данном обзоре).
\end{itemize}

В данной работе мы рассматриваем применение P2P-сетей для создания хранилищ данных. Существуют два больших класса таких хранилищ: 
\begin{itemize}
\item предназначенные для хранения, поиска и обмена публичной информацией (файлообменные системы; file sharing systems);
\begin{itemize}
\item как правило, в таких сетях осуществляется простой одноразовый обмен файлами между компьютерами; существуют средства для поиска и передачи файлов между узлами сети; в типичном случае это ``легковесные'' системы с качеством обслуживания ``по мере возможности'' (best effort), не заботящиеся о безопасности, доступности и живучести;
\end{itemize}
\item предназначенные для хранения приватной информации;
\begin{itemize}
\item такие системы предоставляют среду распределенного хранения, в которой пользователи могут распределять, сохранять и, при желании,  публиковать контент; при этом поддерживаются безопасность и надежность: доступ к контенту контролируется, и узлы должны обладать соответствующими привилегиями для его получения; основными задачами таких систем являются обеспечение безопасности данных и живучести сети, и зачастую их главная цель заключается в создании средств для идентифицируемости, анонимности, а также управления контентом (обновление, удаление, контроль версий).
\end{itemize}
\end{itemize}

Поскольку оба класса хранилищ основаны на P2P-сетях, они обладают рядом общих свойств. В частности, услуги по хранению данных равномерно распределяются между всеми участниками сети, что обеспечивает естественное распределение нагрузки, отсутствие узких мест и выделенных точек отказа системы, а механизмы распределения информации по узлам могут обеспечить безотказность работы системы даже в случае выхода из строя части узлов хранения, то есть обеспечивают стабильность работы в условиях динамически меняющейся сети узлов хранения. Технологии создания хранилищ публичных данных является хорошо развитой областью исследований и практических реализаций. Более того, большой интерес к P2P-сетям в целом в значительной степени связан с большой популярностью файлообменных P2P-сетей, таких как Napster \cite{3}, Gnutella \cite{4}, Kazaa \cite{5}, BitTorrent \cite{6} и др. Эти и подобные системы,  предназначенные для  обмена публичными файлами, продемонстрировали преимущества и инициировали интенсивные исследования P2P-систем в целом (см., например, \cite{7,8,9,10,11}). 

В последние годы P2P-системы проявили себя и как жизнеспособная архитектура для реализации распределенных систем хранения приватных данных. Хотя, как уже указывалось, децентрализованные хранилища публичных и приватных данных основаны на общей базе P2P-сетей, задачи, которые они решают являются существенно разными. Это приводит к тому, что и технологии построения ДЦХ приватных данных принципиально отличаются от технологий, используемых для файлообменных сетей. Отметим, что существует большое количество обзоров, посвященных ДЦХ публичных данных, как популярных (см., например, \cite{12}), так и посвященных глубоким вопросам теории и практики их построения и исследования (см., например, \cite{13,14}). При этом зачастую в обзорах четко не указывается, что они посвящены P2P-сетям для хранения именно публичных данных, что может ввести в заблуждение читателя, не являющегося специалистом в данной области. Технологии построения ДЦХ приватных данных известны существенно меньше, соответствующих обзоров практически не существует. Данная работа призвана по крайней мере частично восполнить этот пробел. Для лучшего понимания общей картины мы рассматриваем оба класса ДЦХ, подчеркивая различия в принципах их построения. Стоит отметить, что в данном обзоре мы не затрагиваем аспекты построения распределенных файловых систем на основе P2P-сетей.

Статья организована следующим образом. В разделе~\ref{sec:DHP} кратко приводятся основные свойства и принципы построения ДЦХ публичных данных. Раздел~\ref{sec:DHPD} посвящен технологиям построения ДЦХ приватных данных. В разделе~\ref{sec:PKR} приведены примеры конкретных разработок ДЦХ публичных и приватных данных. В разделе~\ref{sec:Zak} содержится заключение и общие выводы о достижениях, проблемах и перспективах построения децентрализованых хранилищ.

\vspace{3mm}
\He{ДЕЦЕНТРАЛИЗОВАННЫЕ ХРАНИЛИЩА ПУБЛИЧНЫХ ДАННЫХ: ОСНОВНЫЕ ТЕХНОЛОГИИ ПОСТРОЕНИЯ}
\label{sec:DHP}

Узлы одноранговых (пиринговых -- от английского peer-to-peer (P2P)) компьютерных сетей могут выступать как в качестве клиента, так и в качестве сервера. Децентрализованные  хранилища публичных данных, которые чаще называются файлообменными P2P-сетями, предназначены для совместного использования файлов всеми пользователями данной сети. Другими словами, любой пользователь, используя существующие в данной сети инструменты и методы поиска, может найти на компьютерах других пользователей интересующие его файлы,  выложенные в открытый доступ, и свободно загрузить их на свой компьютер. Для этого пользователь устанавливает на свой компьютер клиентскую программу для работы с конкретной P2P-сетью, и после этого может как предоставлять свои файлы для открытого доступа в сети, так и отправлять запросы на поиск и загрузку нужных ему файлов. Во многих сетях файл может загружаться частями одновременно из нескольких источников; при этом уже загруженные части какого-либо файла, могут сразу служить одним из источников для других пользователей. За счет такого подхода достигается высокая пропускная способность P2P-сетей. Классическим примером является протокол BitTorrent \cite{6}.

Типичная P2P-сеть как правило объединяет компьютеры из административно несвязанных доменов. Участники P2P-сети могут присоединяться или выходить из системы со значительной частотой, так что  P2P-сети являются по своей природе динамическими. Узлы P2P-сети совпадают с узлами интернета и поддерживают информацию о нескольких других узлах (называемых соседями), так что формируется виртуальная оверлейная сеть поверх интернета. Каждая ссылка в оверлейной P2P-сети соответствует последовательности физических ссылок в базовой сети. 

Любая P2P-система хранения данных должна обеспечивать эффективный и отказоустойчивый поиск данных и узлов, на которых они хранятся, а также маршрутизацию запросов и ответов на них. Для выполнения этих требований были разработаны инфраструктуры и алгоритмы различных типов. В соответствии с этим P2P-сети могут быть классифицированы на основе методов контроля за распределением и способов поиска данных, а также по типам топологий оверлейных сетей. 

Заметим, что хотя одноранговость сетей предполагает, что они полностью децентрализованы, на практике этого не всегда придерживаются, и встречаются системы с той или иной степенью централизации. В централизованных P2P-сетях, таких как Napster \cite{3}, центральный реестр ресурсов и некоторая другая информация о сети сохраняется на единственном центральном сервере. Участники сети находят адреса желаемых файлов, запрашивая этот центральный сервер реестров. Такие P2P-сети плохо масштабируются, а центральный сервер реестров является выделенной точкой отказа.  В дальнейшем  мы не будем рассматривать сети такого типа. 

Децентрализованные P2P используют распределенные реестры. Эти системы, в свою очередь, могут быть подразделены на полностью децентрализованные и гибридные сети \cite{15}. Различие между ними заключается в том какие роли играют узлы сети. В полностью децентрализованных системах, таких как Gnutella и Chord, узлы полностью равноправны. В гибридных системах некоторые узлы, которые называются доминирующими \cite{16} или суперпирами \cite{17,18}, обрабатывают поисковые запросы других, обычных пиров. Пиры в P2P-сети часто бывают неоднородными с точки зрения вычислительной мощности, устойчивости и качества интернет-связи. Полностью децентрализованные системы не могут использовать в своих интересах эту неоднородность, в то время как гибридные системы могут. Суперпирам динамически присваивается задача обслуживания небольшой части оверлейной сети с помощью индексирования и кэширования содержащихся в ней файлов.  Суперпиры индексируют файлы, предоставляемые связанными с ними обычными узлами, и в качестве proxy-серверов выполняют поиск от их имени. Поэтому все запросы первоначально направляются к суперпирам. Однако доминирующие узлы или суперпиры должны быть тщательно отобраны, чтобы избежать появления узких мест и выделенных точек отказа. Как правило, это делается автоматически, исходя из их вычислительной мощности и полосы пропускания. Примером гибридной системы является сеть Kazaa \cite{5}.

С точки зрения структуры одноранговые децентрализованные сети делятся на две основные категории: структурированные  и неструктурированные \cite{19,20,21,22}. В структурированных сетях P2P, таких, например, как Chord \cite{23}, точно определены и сетевая архитектура и размещение данных. Системы на основе таких сетей обеспечивают соответствие между данными (скажем, идентификатором файла) и его расположением (например, адресом узла) в форме распределенной хэш-таблицы (Distributed Hash Table, DHT), так что запросы могут быть эффективно направлены узлу с искомым содержанием. Системы на основе таких сетей являются хорошо масштабируемыми. Их недостатком является сложность управления структурой сети, которая обеспечивает эффективную маршрутизацию сообщений в среде с переменным числом узлов.

В неструктурированных сетях P2P, таких как Gnutella \cite{4}, не существует никаких правил, которые определяют местоположение хранящихся данных, а топология сети произвольна. Механизмы поиска могут быть разными -- от самых простых (например, лавинное распространение запросов способами «сначала вширь» (breadth-first) или «сначала вглубь» (depth-first) до тех пор пока искомые данные не будут найдены), до более сложных методов, например, с использованием метода случайного блуждания и индексации маршрутов. Механизмы поиска, применяемые в неструктурированных сетях, имеют очевидное влияние на доступность, надежность и масштабируемость, которую в таких сетях сложнее обеспечить \cite{87}. Зато неструктурированные системы более подходят для сетей с непостоянным числом узлов. 

\begin{figure*}[t]
\begin{center}
\includegraphics[scale=0.5]{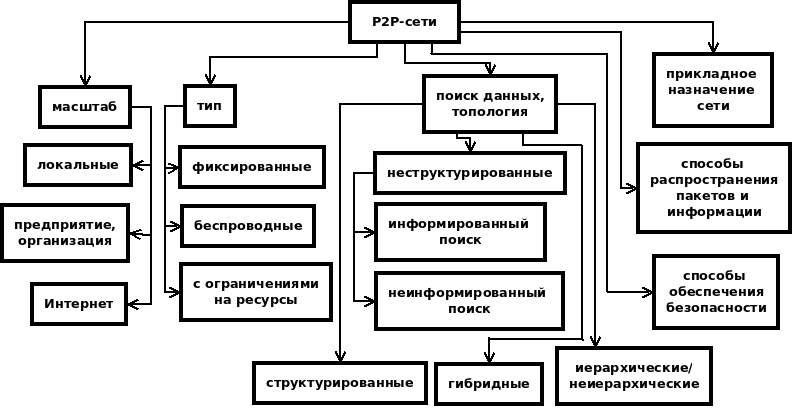}
\end{center}
\caption{Возможная классификация P2P-сетей по различным характеристикам}
\label{fig:fig1}
\end{figure*}

Возможны различные комбинации свойств распределения/поиска данных и свойств оверлейной топологии. Например, в сети Freenet \cite{24} и оверлейная топология, и расположение/поиск данных осуществляются вероятностным образом на основе некоторых предположений; в сети Symphony \cite{25} оверлейная топология определена вероятностным образом, а расположение данных определено точно. Сети такого типа, занимающие промежуточное положение между структурированными и неструктурированными, зачастую называют слабоструктурированными. Существуют и другие характеристики, на основе которых можно классифицировать P2P-системы. В частности, они могут быть подразделены на иерархические и неиерархические в зависимости от того, является ли оверлейная сеть иерархией или нет. Большинство полностью децентрализованных систем имеют плоские оверлейные сети и являются неиерархическими системами. Все гибридные системы и некоторые полностью децентрализованные системы, таких как Kelips \cite{26}, являются иерархическими. Неиерархические системы позволяют обеспечить  балансировку нагрузки и высокую устойчивость. Иерархические системы обеспечивают хорошую масштабируемость, возможность использовать неоднородность узлов для повышения эффективности работы сети и оптимальную маршрутизацию. Существуют также многослойные оверлейные сети \cite{27,28,29}. На рисунке~\ref{fig:fig1} показана возможная классификация \cite{14} децентрализованных P2P-сетей по различным характеристикам.

Структура  оверлейных P2P-сетей является предметом интенсивных исследований в последние годы (см., например,  \cite{13} и ссылки в нем). В структурированных оверлейных сетях узлы совместно поддерживают маршрутную информацию о том, как достигнуть любого узла \cite{30,31}. По сравнению с неструктурированными сетями структурированные оверлейные сети обеспечивают существование верхнего предела для количества сообщений, необходимого для нахождения любого объекта в сети. Это особенно важно при поиске объектов, которые запрашиваются редко. Чтобы обеспечить детерминистскую маршрутизацию, пиры размещены в виртуальном адресном пространстве, оверлейная сеть соответствует определенной геометрии, и в объединенном пространстве объектов и идентификаторов узлов определена метрика, на которой базируется алгоритм маршрутизации сообщений. У каждого пира есть локальная таблица маршрутизации, которая используется передающим алгоритмом. Таблица маршрутизации пира инициализируется, когда он присоединяется к сети с использованием специальной процедуры начальной загрузки. Пиры периодически обмениваются информацией об изменениях таблицы маршрутизации. 
В большинстве случаев структурированных оверлейных сетей узлы и ресурсы совместно используют одно и то же пространство идентификаторов и  отображаются на это пространство посредством согласованных хеш-функций. При этом используется основанная на ключах маршрутизация, при которой  пир с данным идентификатором хранит  ресурсы с ключами, близкими его идентификатору в смысле указанного выше общего пространства. Для реализации соответствующего алгоритма часто используются распределенные хэш-таблицы (Distributed Hash Tables; DHT) \cite{32}. Примерами систем, использующих структурированные оверлейные сети являются: Chord \cite{23}; CAN \cite{33}; Pastry \cite{34}; Tapestry \cite{34}; Kademlia \cite{36}; Viceroy \cite{37}; P-Grid \cite{38}; SkipNet \cite{39}. 
Графы неструктурированных оверлейных сетей, как правило, имеют структуру совпадающую или близкую по свойствам  случайным графам, безмасштабным (со степенным распределением степеней узлов) графам, или графам малого мира \cite{40,41,42}.  Важной частью исследований неструктурированных сетей является построение оптимальных свойств графа для них и разработка децентрализованных алгоритмов формирования и поддержки этих свойств и характеристик при динамически изменяющемся составе пиров и хранящихся данных \cite{43,44,45}. Примерами систем, использующих неструктурированные оверлейные сети являются: Freenet \cite{24}; Gnutella \cite{4}; FastTrack \cite{46}; BitTorrent \cite{47}; UMM \cite{48}; Gia \cite{49}; Phenix \cite{50}. 
Особую группу неструктурированных P2P-сетей, которые активно разрабатываются и исследуются в последнее время, представляют так называемые биотехнологические сети или, иначе говоря, сети с агент-ориентированной архитектурой \cite{51}. Биотехнологические решения характеризуются высокой степенью адаптивности и реактивности поведения, внутренней поддержкой неоднородности сети, естественно-распределенным характером функционирования, устойчивостью к отказу компонентов и возможностью самоорганизации \cite{52,53,54}. В настоящее время становится очевидным, что такой подход является очень перспективным кандидатом на то, чтобы решить проблемы динамического изменения P2P-сети и управления топологией оверлейной сети \cite{55}. Решения на основе стайного интеллекта, а именно, на основе коллективного поведения колонии муравьев или роя пчел, обеспечивают масштабируемость из-за распределенного интеллекта и уменьшенных затрат на коммуникацию, что достигается посредством механизма стигмергии (англ. - stigmergy), то есть спонтанного непрямого (через среду обитания) взаимодействия между индивидами стаи. Примерами систем, использующих биотехнологии для разработки агент-ориентированной архитектуры P2P-сетей являются: BlantAnt \cite{56}; AntOM \cite{57}; Self-Chord \cite{58}. 
Большое внимание при исследованиях и разработках P2P оверлейных сетей, особенно неструктурированных, включая биотехнологические, уделялось разработке алгоритмов эффективного размещения и поиска данных, а также распространения и обработки запросов на их получение. К настоящему времени разработано множество таких алгоритмов и их модификаций и адаптаций для различных оверлейных топологий и конкретных реализаций сетей  \cite{13,14}. Общей основой для анализа оверлейных топологий и алгоритмов поиска данных является теория сложных сетей \cite{40,41}. В частности, во многих случаях разработчики стремятся получить оверлейные сети со свойствами «малого мира», одним из важнейших свойств которых является малое среднее расстояние между узлами и малый диаметр сети (диаметр сети -- это наибольшее расстояние между узлами). Более точное выражение этого свойства заключается в следующем: для регулярной решетки среднее расстояние между узлами растет как степень числа узлов, а для сети со свойствами малого мира существенно медленнее -- логарифмически. Необходимо отметить, что помимо малой средней длины пути между узлами еще одним общим отличительным свойством сетей со свойствами малого мира является высокая степень кластеризации. Высокая кластеризация обеспечивает локальную устойчивость сети: существование локальных обходных путей при выходе из строя какого-либо узла сети. В рамках теории сложных сетей рассматривают не только статистические, но динамические сети, для описания структуры которых необходимо учитывать принципы их эволюции, что важно для понимания механизма образования оверлейных P2P-сетей различных типов. При разработке архитектуры оверлейной P2P-сети необходимо учитывать, что ее узлы совпадают с узлами физической сети Интернет. Поэтому топологические аспекты оверлейной  сети будут неизбежно коррелировать со свойствами физической сети. В частности может оказаться, что построенная оверлейная сеть плохо согласуется с базовой физической (так называемая, проблема ``topology mismatching''). Такие взаимосвязи изучаются в теории пространственно-вложенных сложных сетей (см., например, обзор \cite{59} и ссылки в нем). 
Cуществующие файлообменные сети практически не предоставляют средств обеспечения безопасности и надежности. Эта проблема давно известна (см., например, \cite{60,61,62}), однако разработки в данном направлении развиты весьма слабо и в настоящее время. Обеспечение надежности, чаще всего, сводится к простой репликации, причем зачастую репликация является стихийной, неуправляемой (например, за счет того, что многие участники сети  держат на своем ресурсе какой-либо фильм или электронную книгу поскольку они им просто нравятся). В случае сети BitTorrent репликация сопровождается фрагментированием файла, однако это не имеет прямого отношения к надежности доступа (так как если участник сети имеет данный контент, то он имеет все его фрагменты), а предназначено для увеличения скорости передачи информации. 

\He{ДЕЦЕНТРАЛИЗОВАННЫЕ ХРАНИЛИЩА ПРИВАТНЫХ ДАННЫХ: ОСНОВНЫЕ ТЕХНОЛОГИИ ПОСТРОЕНИЯ}
\label{sec:DHPD}

Основными задачами, которые должны решать децентрализованных хранилищ приватных данных являются:
\begin{enumerate}
\item безопасное хранение данных на географически разделенных ресурсах интернет-пользователей с обеспечением приватности (изолированности) информации и гибкого управления правами доступа к ней;
\item обеспечение надежности и устойчивости хранения, что подразумевает сохранность (самовосстановление) информации при всевозможных ошибках и сбоях в системе;
\item обеспечение постоянной доступности информации, что означает устойчивость системы в целом при выходе из строя некоторого ее сегмента;
\item предоставление удобного доступа пользователей к данным, контроль и учет хранения данных, причем средства контроля и учета также должны обладать высокой степенью надежности (в частности, высокой степенью защиты от незаконного проникновения и и изменения контрольно-учетной информации), а также отказоустойчивости;
\item возможность масштабирования системы хранения и наращивания (в случае необходимости) ее функциональных возможностей без перестройки всей системы.
\end{enumerate}

Эти задачи сходны с теми, которые стоят и перед распределенными системами хранения данных с клиент-серверной архитектурой \cite{85,86}, однако решаются они существенно другими способами. Первая задача решается на основе использования алгоритма рассредоточения информации (Information Dispersal Algorithms, IDA, \cite{63}), благодаря которому каждый владелец ресурса хранения локально имеет доступ только к закодированному фрагменту единицы информации (файла), что принципиально не позволяет этому владельцу восстановить даже часть информации. Кроме того, управление правами доступа к информации на уровне всей системы может осуществляться путем создания инфраструктуры безопасности на основе электронных мандатов для пользователей системы.

Цели задач 2 и 3 достигаются благодаря тому, что система хранения является децентрализованной и географически распределенной. Это существенно снижает риск отказа или даже физического уничтожения (в случае чрезвычайных обстоятельств, например, пожара и т.п.) всей системы в целом. Кроме того, важно, что в рамках IDA данные кодируются избыточным образом, делятся на фрагменты, которые распределяется (посредством защищенных интернет-соединений) по разным географически распределенным узлам хранения. Данные в такой системе можно восстановить, даже если они будут уничтожены в части пунктов хранения. Как уже отмечалось, индивидуальные фрагменты, однако, не несут достаточной информации для восстановления всех и даже части данных из исходного файла, чем обеспечивается, наряду с системой управления правами доступа, приватность хранимой информации. 

Задачи 4 и 5 решаются с помощью средств удаленного (защищенного) доступа к данным с удобными интерфейсами пользователя, а также использования современных подходов к построению распределенных систем, в частности микросервисной архитектуры и модульного подхода с четко определенными API для каждого компонента системы, что обеспечивает возможность неограниченного расширения ресурсов хранения интернет пользователей, подключенных к системе. Удобство (скорость доступа) и масштабируемость системы также существенно зависят от правильно выбранных или разработанных протоколов передачи информации. 

Важно подчеркнуть, что одна из основных проблем, присущих публичным ДЦХ, а именно обеспечение эффективного поиска и маршрутизации требуемых данных, в случае хранилищ приватных данных в значительной степени теряет свою актуальность -- пользователь, который размещает на ресурсах сети свои данные просто хранит информацию об их локализации (адресах) и отслеживает возможные перемещения своих данных между узлами сети. Это информация может храниться на локальных ресурсах пользователя, или в распределенной базе данных -- важно, что она доступна только владельцу данных, или группе пользователей (владельцев), которые имеют доступ к этой информации и права на совершение тех или иных действий с данными (чтение, модификация, удаление).

Вместе с тем, как указано выше, существуют принципиально новые по сравнению с публичными ДЦХ задачи -- обеспечение конфиденциальности данных, надежности хранения (включая обеспечение постоянной уверенности пользователя в том, что его данные действительно хранятся в неискаженном виде на удаленных распределенных ресурсах). Решение этих задач, а также создание материальных стимулов для предоставления интернет-пользователями части своих ресурсов хранения, без чего создание P2P-сети хранения приватных данных принципиально невозможно, обеспечивается технологиями доверенного распределенного учета потребления ресурсов, инструментария для автоматизированного заключения контрактов между потребителями и провайдерами ресурсов, а также системами взаиморасчетов (формирования счетов).

Ниже мы подробнее остановимся на этих специфических аспектах построения ДЦХ для приватных данных.

Простейший вариант решения задачи обеспечения надежности и приватности хранения -- это шифрование файлов и хранение нескольких копий зашифрованных файлов на различных носителях. Однако такой вариант решения задачи является самым затратным по объемам хранения и передачи данных и избыточным по надежности. Поэтому для надежного  хранения файлов и обеспечения приватности хранимой информации   используется алгоритм рассредоточения информации (Information Dispersal Algorithms -- IDA) \cite{63,64}. Этот метод в разы повышает надежность  по сравнению с обычной репликацией, которая чаще всего используется в одноранговых хранилищах для публичных данных \cite{65}. Очень важно, что каждая из распределенных частей информации в результате работы IDA принципиально не поддается расшифровке. Это обеспечивает кофиденциальность хранения приватного контента на публичных ресурсах. При этом существуют различные конкретные варианты IDA, которые можно выбирать из соображений оптимизации использования сетевых, вычислительных ресурсов и ресурсов хранения данных (см. обзор \cite{64} и ссылки в нем). 

В соответствии с IDA информация, хранящаяся в файле, кодируется с одновременным разделением на n фрагментов меньшего размера таким образом, чтобы исходный файл можно было восстановить по любым $k < n$ фрагментам, а общий объем хранения фрагментов был  меньше, чем при простой репликации. С точки зрения общей теории криптографии, IDA  принадлежит классу алгоритмов разделения секрета. Эти алгоритмы позволяют разделить какой-либо секрет между несколькими участниками группы так, что каждому известна только его часть, а весь секрет может быть восстановлен только некоторой подгруппой участников первоначальной группы, причем в эту подгруппу должно входить не менее некоторого изначально известного числа участников. Алгоритмы разделения секрета применяются в случаях, когда существует значимая вероятность компрометации одного или нескольких хранителей секрета, но вероятность недобросовестного сговора значительной части участников считается пренебрежимо малой.

С точки зрения ДЦХ для приватных данных провайдер, предоставивший свои ресурсы  хранения, имеет лишь закодированный фрагмент данных (часть секрета), но даже не знает где хранятся другие фрагменты, необходимые для того, чтобы была принципиальная возможность расшифровки хранимой информации, не говоря уже о сложностях собственно расшифровки и нахождения достаточной мотивации для того, чтобы другие участник сети согласились на противоправные действия (предоставили свои фрагменты стороннему лицу).

Наиболее известный способ кодирования в рамках IDA основан на кодах Рида-Соломона \cite{66}, которые базируются на блочном принципе коррекции ошибок и используются в огромном числе приложений в сфере цифровых телекоммуникаций и при построении запоминающих устройств. Однако, при использовании IDA в рамках P2P-сети коды  Рида-Соломона могут оказаться слишком тяжеловесными с точки зрения потребления вычислительных и сетевых ресурсов. При потере, например из-за выхода из строя части узлов сети, некоторого числа фрагментов исходного файла (но не более некоторого фиксированного числа $n-k$, являющегося параметром алгоритма) возможно его восстановление, но для этого необходимо выполнить вычислительные операции по восстановлению данных и передачу восстановленных частей на другие -- работоспособные -- узлы. Поэтому, как уже упоминалось, в настоящее время интенсивно исследуются различные другие версии IDA, позволяющие оптимизировать алгоритм и используемые коды и сбалансировать требования надежности и потребление ресурсов сети \cite{64}.

Дополнительным преимуществом при использовании IDA является является возможность параллельной передачи фрагментов файла с пропорциональным уменьшением времени. До некоторой степени это аналогично сети BitTorrent (для публичных данных), в которой обычная репликация сопровождается фрагментированием файла, однако в случае BitTorrent  это не имеет прямого отношения к надежности доступа (так как если участник сети имеет данный контент, то он имеет все его фрагменты), а предназначено именно для увеличения скорости передачи информации. 

Еще одним ключевым аспектом информационной безопасности в рамках ДЦХ для приватных данных является целостность хранимой информации. Целостность означает, что данные могут быть изменены только уполномоченными сторонами или уполномоченными способами. Это относится к защите данных от несанкционированного удаления или изменения. Другими словами данные должны честно храниться на ресурсах P2P-сети (заметим, что это актуально и для облачных хранилищ), а любые нарушения, то есть, потеря, изменение или компрометация данных, должны быть обнаружены. 

Поскольку пользователи в случае ДЦХ локально не обладают своими данными, они не могут использовать традиционные криптографические способы для защиты их сохранности и неизменности. Для того, чтобы пользователь мог убедиться, что его данные действительно хранятся на удаленных ресурсах без необходимости полностью загружать их на локальный компьютер (что неприемлемо с точки зрения величины накладных расходов), были разработаны специальные алгоритмы -- доказуемое хранение данных (Provable Data Possession; PDP) \cite{67,68,69} и доказательство возможности извлечения (Proof of Retrievability; POR) \cite{70,71} (см. также обзор \cite{72} и ссылки в нем). Эти алгоритмы являются аналогами хорошо известных в криптографии доказательств с нулевым разглашением информации  (Zero-knowledge proof) \cite{73}, то есть интерактивных криптографических протоколов, позволяющих одной из взаимодействующих сторон убедить другую в том, что она обладает некоторой информаций, не раскрывая ее содержание. 

PDP-алгоритм является более легким в реализации. Его основная идея состоит в том, что пользователь вычисляет значения хеш-функции $H(K, F)$ для файла $F$ с ключом $K$, а затем отправляет $F$ в хранилище. Когда пользователь считает нужным проверить сохранность файла, он посылает ключ $K$ на узел хранения и просит пересчитать хэш-значение, основанное на $F$ и $K$. После этого узел возвращает хеш-результат пользователю для сравнения. Пользователь может инициировать большое число проверок, сохраняя набор ключей и соответствующих значений хэш-функции. Такой подход обеспечивает достаточно строгое доказательство того, что сервер все еще сохраняет $F$. Таким образом, эта схема работает по типу задание-ответ (challenge and response protocol). Обычно PDP применимо только к статически хранящимся файлам (архивам).

POR также является достаточно легким  протоколом с точки зрения потребления ресурсов. Пользователь хранит только ключ, который используется для кодирования файла $F$,  в результате которого получается зашифрованный файл $F^\prime$. Основная идея состоит в том, что в $F^\prime$ встраивается набор сигнальных меток, которые в $F^\prime$ неотличимы от содержательных обычных блоков данных, и поэтому владелец ресурсов хранения на узле сети заранее не может определить их местоположение. В рамках протокола задание-ответ, владельца узла  просят вернуть определенное подмножество меток из $F^\prime$. Если файл $F^\prime$ удален или подделан существует высокая вероятность того, что по крайней мере часть меток тоже удалена или повреждена, что приводит к тому, что узел оказывается не в состоянии доказать наличие неискаженного файла. Если же ответ правильный, владелец с высокой вероятностью может быть уверен, что его данные хранятся в неизменном виде.

Существуют усложненные версии PDP  и POR, которые, в частности, обладают лучшей масштабируемостью и возможностью проверять корректность динамически изменяемых файлов \cite{72}.

Очевидно, что для того, чтобы ДЦХ для приватных данных на основе P2P-сетей могло реально работать, пиры, которые предоставляют свои ресурсы для хранения, должны получать за это вознаграждение. Поэтому система децентрализованного хранения должна обеспечивать формирование контрактов на хранение между пирами в духе парадигмы так называемых ``умных контрактов'' \cite{74}. В данном случае контракты являются  соглашениями между поставщиком услуг по хранению данных и их потребителем, которые определяют -- какие данные будут храниться и какова стоимость хранения. Они также включают требование, чтобы поставщик услуг по хранению данных постоянно был готов доказать (например, с помощью протоколов PDP или POR, описанных выше), что данные хранятся в неизменном виде. Для того, чтобы сохранить принцип полной децентрализации контракты должны храниться в распределенном реестре. Наиболее подходящей для создания таких реестров является технология на основе блокчейнов (blockchain) \cite{75}, которая обеспечивает их публичный аудит. Одной из интенсивно развиваемых в последнее время платформ для создания децентрализованных приложений с использованием технологии блокчейна является Ethereum \cite{76,77}. Ethereum использует основные наработки, предложенные в рамках криптовалюты  Bitcoin \cite{78}, протокола BitTorrent и идею умных контрактов для создания общей платформы, позволяющей разработчикам использовать эти новые технологии для различных целей. Ethereum позволяет оперировать собственной криптовалютой (эфир), что позволяет обеспечить взаиморасчеты между участниками контрактов за предоставленные услуги. Поскольку Ethereum явлется блокчейн-платформой общего назначения, мы ее здесь подробнее не рассматриваем. Две другие -- специализированные для ДЦХ приватных данных -- платформы (Storj и Sia) представлены в следующем разделе.

\He{ПРИМЕРЫ КОНКРЕТНЫХ РАЗРАБОТОК}
\label{sec:PKR}
\he{Децентрализованные хранилища публичных данных}
\label{sec:PKR-PUB}

\textit{Chord и Self-Chord}

P2P-сеть Chord \cite{23} основана на технологии распределенных хеш-таблиц (Distributed Hash Table, DHT) \cite{32} (другие известные P2P-сети на основе DHT -- CAN \cite{33}, Tapestry and Pastry \cite{34}). Пространство ключей хеш-таблицы Chord представляется в виде окружности. Все узлы сети имеют идентификаторы, которые принимают значения в том же пространстве, что и ключи. Идентификаторы и ключи получаются в результате так называемого консистентного хеширования. Консистентное хеширование позволяет равномерно распределять ключи и идентификаторы узлов на пространстве ключей хеш-таблицы, а также позволяет узлам присоединяться и покидать систему без прерывания работы системы. Для узла идентификатором является хеш-значение от его IP адреса, для ключа идентификатор определяется как хэш-значение от какого-либо ключевого слова, например от имени файла.

В случае $m$-битных идентификаторов и ключей  максимальное количество узлов не может превышать $2m$, так что окружность  содержит ключи и идентификаторы от $0$ до $(2m-1)$. Каждый узел имеет потомка и предшественника: потомком узла (или ключа) является следующий по ходу часовой стрелки узел (ключ) на окружности значений, а предшественником является предыдущий. Так как потомок (или предшественник) может покинуть сеть (по причине отказа или корректного выхода), каждый узел хранит информацию о некотором сегменте круга, примыкающего к нему. Результатом является высокая вероятность того, что узел способен корректно определить потомка или предшественника, даже если в сети высока интенсивность отказов узлов. Если имеется $N$ узлов и $K$ ключей, то каждый узел при равномерном распределения идентификаторов узлов на окружности отвечает примерно за $K/N$ ключей. В случае присоединения нового узла или отказа/отключения существовавшего узла, изменяется ответственность за ${\cal O}(K/N)$ ключей. Если каждый узел знает только позицию своего потомка, то последовательный поиск по сети позволяет найти запрашиваемый ключ, но это самый простой и медленный метод поиска, так как может потребоваться, чтобы запрос был передан через большой сегмент сети. Поэтому протокол Chord реализует более быстрый метод поиска: каждый узел имеет таблицу, содержащую записи о $m$ узлах таким образом, что $i$-я запись узла $n$ содержит адрес узла \{потомок $((n + 2i-1)$ mod $2m$ \}. Это делает топологию оверлейной сети подобной топологии ``малого мира''. Поэтому неудивительно, что с такой таблицей маршрутизации количество узлов, с которыми необходимо будет связываться, чтобы найти нужный узел, для сети из $N$ узлов является ${\cal O}(log(N))$, то есть растет логарифмически.

Cуществует целый ряд других систем, использующих топологию кольца. Например, в работе \cite{79} изучались различные методы оптимизации для систем на основе DHT, таких как стратегии репликации, использования алгоритмов удаляющего кодирования (erasure coding),  итеративной и рекурсивной маршрутизации. Оптимизированная система на основе Chord, названная DHash++ , является результатом этих исследований. Система Hybrid-Chord, разработанная в работе \cite{80}, повышает производительность и надежность Chord путем введения некоторой избыточности в систему с помощью нескольких Chord-колец поверх друг друга, и с помощью нескольких списков потомков постоянного размера.

Весьма интересное и перспективное развитие Chord-подобного протокола связано с  использованием биотехнологии для разработки агент-ориентированной и сомоорганизующейся системы, названной Self-Chord \cite{58}. В Self-Chord пространства ключей и идентификаторов узлов могут быть разными, причем ключи распределяются по кольцу на  основе статистического подхода и парадигмы стайного интеллекта. Такое распределение гарантирует, что похожие ключи хранятся на соседних узлах и что процедура поиска, включая запросы по диапазону, могут быть выполнены за время, логарифмически зависящее от размера сети. Распределение (или перераспределение -- при появлении новых ключей и появлении/исчезновении узлов) осуществляется независимыми мобильными агентами, которые перемещаются по оверлейной сети и коллективно перестраивают идентификаторы ресурсов, чтобы способствовать эффективности поиска.

Важное отличие Self-Chord от Chord заключается в том, что в то время как в последней ключ помещаются в узел, идентификатор которого близок по значению этому узлу, в Self-Chord, ресурсы помещаются в узлы на основе критериев выравнивания нагрузки, а также так, что семантически близкие ресурсы помещаются на соседних узлах (кластеризуются).

Это позволяет пользователям легко выполнять запросы, критерии которых определяются некоторым диапазоном их значений, так как достаточно найти определенный информационный ресурс и затем получить доступ к ресурсам, которые расположены на том же узле, чтобы гарантировать, что пользователь найдет другие семантически близкие ресурсы. Кроме того, управление системами на основке Self-Chord  является более простым, чем в случае Chord, так как нет необходимости частой перестройки идентификаторов ресурсов из-за появления и исчезновения узлов (churn-проблема) благодаря непрерывной оптимизации, выполняемой сетевыми агентами.

Важно отметить, что, технология, разработанная для Self-Chord может быть распространена на другие структурированные сети, такие, например, как CAN \cite{81}.

\vspace{3mm}
\textit{Freenet}

Cеть Freenet \cite{24} позволяет хранить данные и извлекать их при помощи связанного с ними ключа, причем обладает высокой живучестью при полной анонимности и децентрализации всех процессов в сети. Является типичным примером полностью децентрализованной неструктурированной (иногда ее характеризуют как слабоструктурированную) самоорганизующейся системы распределения данных по ресурсам пользователей одноранговой сети. 

Файлы в рамках Freenet идентифицируются с помощью двоичных ключей, простейший алгоритм получения которых состоит в применении хэш-функции к короткой строке метаданных о файле. Чтобы найти файл, пользователь посылает запрос, содержащий ключ и время жизни запроса (Time to Live; TTL), выраженное через максимально допустимое количество пройденных транзитных узлов (Hops to Live). Для присоединения к Freenet новый пользователь прежде всего определяет адрес одного или более уже существующих узлов. Чтобы поместить новый файл в сеть, узел сначала его шифрует, при желании, делит его на фрагменты (реализуя простой вариант IDA), вычисляет его двоичный ключ и посылает запрос на сохранение файла (Data insert) известным узлам. Важно, что этот  запрос не имеет конкретного указания -- на каком узле желательно сохранить файл: он может быть выполнен как первом узле, куда пришел запрос,  так передан другому соседу. При этом узел, получивший такое сообщение, проверяет -- не  используется ли уже этот ключ. Если ключ уже используется, узел возвращает существующий файл источнику. Если ключ оказывается уникальным, узел использует свою маршрутную таблицу для отправки полученного сообщения с данными ближайшему в смысле лексикографической метрики узлу. Это продолжается до тех пор, пока не не будет исчерпан параметр Hops to Live. В результате работы этого алгоритма новые файлы размещаются на нескольких узлах, уже владеющих файлами с подобными ключами. Одновременно все узлы, участвовавшие в передаче сообщения и данных, обновляют свои маршрутные таблицы, так что новый узел становится известным другим узлам в сети. Если весь процесс прошел успешно и в результате не обнаружилось коллизии значений ключей, подтверждение об успешном размещении данных в сети направляется источнику. Данные в хранилище нельзя редактировать. Пока они находятся в хранилище, они всегда связаны с одним ключом. При этом поместивший данный файл на хранение узел фактически не знает от кого он пришел: если параметр  Hops to Live имеет значение $m$, а узлы сети имеют в среднем $k$ соседей, то существует $k^m$ вариантов возможного отправителя; например, при $m=10$ и $k=3$ получается гигантское число, и фактически любой узел сети может оказаться отправителем. 

Упрощенный алгоритм поиска данных состоит в следующем. Если запрос приходит на узел, где хранится запрашиваемый файл, поиск останавливается и данные направляются  инициатору запроса. Если требуемого файла нет, узел передает запрос соседнему узлу, который, в соответствии с маршрутной таблицей текущего узла, хранит ключ, наиболее близкий к требуемому. Также как и при записи невозможно определить -- какому узлу реально потребовался хранящийся файл или его фрагмент -- соседу или весьма удаленному узлу. Так обеспечивается анонимность в сети Freenet. 

Алгоритмы распределения и поиска файлов обеспечивают накапливание информации о расположении похожих файлов в рамках оверлейной сети Freenet и, соответственно, уменьшение времени отклика из-за уменьшения числа скачков между узлами для достижения цели. А сопровождающий их работу процесс репликации увеличивает устойчивость работы сети при отказе узлов. Кроме того, между источниками запросов и узлом-ответчиком устанавливаются прямые связи, тем самым постоянно изменяя оверлейную сеть и уменьшая в будущем среднее время поиска в сети. Подобный механизм используется и для размещения данных, причем существуют механизмы удаления устаревших записей в таблицах маршрутизации и в хранилищах данных.

Результаты исследований \cite{82} показывают, что средняя длина пути сообщения при поиске в процессе работы сети быстро падает (например для сети из 1000 узлов до приблизительно 6 скачков), что доказывает преимущества механизмов репликации. С точки зрения масштабируемости моделирование показало, что эта средняя длина пути растет приблизительно логарифмическим образом, причем не слишком сильно растет и при отказе части (до 30\%) узлов. Эта отказоустойчивость объясняется сходством топологии оверлейной сети Freenet с топологией «малого мира».

В сети Freenet существует примитивная система поиска по ключевым словам благодаря тому, что несколько сайтов сети содержат перечень ресурсов, опубликованных на остальных сайтах Freenet -- директорию. Это, конечно, нарушает принцип полной децентрализации и является возможным в силу относительно небольшого размера сети. Во время создания нового сайта автор может добавить свой сайт в этот перечень, тем самым позволяя другим пользователям обнаружить его. Владелец директории также периодически запускает робота, который проверяет сайты директории на наличие в них ссылок на сторонние сайты Freenet. Одной из наиболее известных директорий является Freedom Engine. Другими словами, поиск информации действует по принципу ``сарафанного радио'': каждый автор дает на своей странице ссылки на заинтересовавшие его ресурсы, а другой пользователь, которому понравился этот автор, получает, в свою очередь, подборку ссылок по интересам.

Для того чтобы работать в сети Freenet, необходимо установить на компьютер одноименное программное обеспечение, после чего в браузере открывается страница с пошаговыми инструкциями по настройке безопасности. Существуют также специализированные приложения, работающие в сети Freenet, в частности Frost -- система форумов в сети Freenet. Другие программы обеспечивают выполнение других различных вспомогательных программ, например jSite позволяет легко создавать свои собственные фрисайты, Thaw -- работать  с группами файлов, freemulet -- обеспечивать удобный обмен файлами.

\vspace{16mm}
\he{Децентрализованные хранилища приватных данных}
\label{sec:PKR-PRIV}

\textit{Storj}

Облачная сеть хранения данных Storj (storj.io) \cite{83}, функционирующая в тестовом режиме, дает пользователям возможность предоставлять в аренду место на жестких дисках своих компьютеров, не требуя для этого каких-то специфических знаний в области информационных технологий, благодаря тому, среди прочего, что существует графический интерфейс к сервисам сети. Пользователи, предоставляющие место для хранения, в перспективе смогут получить за это награду в криптовалюте Storjcoin X (SJCX) причем размер награды будет зависеть от предоставленного объема и срока эксплуатации. Криптовалюта SJCX стала доступной для обращения с июля 2014 года, после того, как компания Storj устроила распродажу своих токенов с целью финансирования платформы. Монеты были включены в валютные списки нескольких бирж, в том числе Poloniex (poloniex.com). 

Система Storj в полной мере использует технологии, изложенные в разделе \ref{sec:DHPD}, в частности:
\begin{itemize}
\item использование IDA для обеспечения безопасности и приватности данных;
\item поддержка целостности данных с помощью некоторого варианта POR-алгоритма;
\item использование технологии блокчейн и криптовалюты для расчетов между потребителями и поставщиками ресурсов хранения;
\item инфраструктура сети обеспечивает защиту от сбоев и несанкционированного доступа. 
\end{itemize}

Базовый протокол, который формирует распределенную систему, помогает узлам заключать контракты, облегчая передачу данных, проверку целостности и доступности удаленных данных, извлечения данных, а также плату за хранение данных на узлах сети. Каждый одноранговый узел является самоуправляемым агентом, значительно уменьшая человеческое взаимодействие в этом процессе. В системе используется концепция микроплатежей и позволяет сочетать платежи с аудитом. При этом  механизм микроплатежей приводит к ограничению скорости изменения курса валюты, этому способствует и то, что единственная цель Storjcoin -- использование в сети Storj. В результате Storj заработал доверие со стороны рынка: к концу 2016 года создатели Storjcoin объявили его одной из  лучших криптовалют по прибыльности (увеличение цены за год примерно на 700\%).

Сеть Storj уже сейчас предлагает порядка десяти петабайт для хранения данных предоставляемых несколькими сотнями «фермеров» (провайдеры ресурсов хранения в терминологии Storj). Сеть вступила в партнерство с Microsoft Azure в качестве участника программы «Блокчейн-как-услуга» (BaaS, Blockchain-as-a-Service; BaaS). Корпоративные клиенты платформы Azure смогут пользоваться услугами Storj через свой аккаунт в Azure. Как указано в блоге Storj, интерес к Storj уже проявили такие компании, как Coca-Cola, Cox Enterprises (которая намерена платить за загрузку одного петабайта в сеть), InterContinental Hotels Group, The Weather Channel и Capgemini. Они в настоящее время изучают возможности коммерциализации Storj в рамках своих организаций.  

У проекта Storj имеется несколько конкурентов, среди наиболее известных -- платформа Ethereum \cite{76,77}, упоминавшаяся в разделе \ref{sec:DHPD} и Sia (см. ниже). 

\vspace{3mm}
\textit{Sia}

Проект Sia (sia.tech) \cite{84} также участвует в разработке и внедрении децентрализованной облачной инфраструктуры хранения данных. Как и в случае со Storj, сеть Sia будет оперировать собственной криптовалютой -- Siacoin, необходимой для покупки контрактов на определенные объемы дискового пространства.  На данный момент Sia уже представляет собой рабочую сеть, где одни пользователи предоставляют свободные ресурсы хранения, а другие используют их.  

Sia обладает всеми функциональными возможностями, которые необходимы для работы с сервисом, включая опции создания смарт-контрактов, загрузки файлов и выделения свободного дискового пространства в сеть. Когда пользователь хочет предоставить свои ресурсы хранения, он устанавливает собственную цену в SiaCoin за ГБ/месяц и указывает объем дискового пространства, который можете выделить в сеть. Затем это предложение выставляется на биржу, и пользователь ждет когда на это предложение найдется спрос. Когда появляются желающие купить ресурсы, заключаются контракты на хранение (storage contracts), которые хранятся и валидируются на основе блокчейн-технологии. При загрузке файла в Sia формируется контракт на его хранение, содержащий специальный хэш файла (Merkle root), с вознаграждением или штрафом для хоста, которые выплачиваются в Siacoins в момент последующей проверки файла на целостность (в духе POR). Надежность и приватность в Sia достигается благодаря использованию варианта IDA, причем авторы Sia, применяя статистический анализ, ожидают показатель надежности данных на уровне 99,9999\%.

В будущем Sia надеется стать децентрализованной сетью дата-центров и ресурсов хранения индивидуальных пользователей Интернета, которые, объединенные вместе платформой Sia, создадут крупную,  быструю, дешевую и безопасную сеть для хранения данных. Исходный код Sia опубликован на GitHub под свободной лицензией MIT.

\He{ЗАКЛЮЧЕНИЕ}
\label{sec:Zak}
В P2P-сетях конечные пользователи совместно используют ресурсы через прямой обмен между компьютерами. Информация распределена между участвующими в сети узлами вместо концентрации на единственном или нескольких серверах. Чистая одноранговая система это распределенная система без централизованного управления, где программное обеспечение, работающее в каждом узле, является по своей функциональности одинаковым. Таким образом, узлы в P2P-сети обычно играют равнозначные роли, поэтому, эти узлы также называют пирами (от английского peer - ровня). Привлекательные свойства одноранговой архитектуры породили множество научно-исследовательских работ в области создания распределенных файловых P2P-систем. Полученные результаты и достигнутые успехи в этой области обуславливают то, что P2P-системы являются одним из главных предметов текущих и будущих научных исследований в области распределенных систем хранения. 

Важно однако осознавать существенную разницу между децентрализованными хранилищами двух классов: предназначенных для хранения, поиска и обмена публичной информацией (файлообменные системы) и предназначенных для хранения приватной информации. В данной работе продемонстрировано, что хотя оба класса хранилищ основаны на парадигме децентрализации и одноранговых (P2P) сетях, и в силу этого обладают рядом общих свойств, однако технологии их создания принципиально разные. 

Развитие технологий создания хранилищ публичных данных имеют существенно более длинную историю и является хорошо развитой областью исследований и практических реализаций. Жизнеспособные реализации распределенных систем хранения приватных данных появились сравнительно недавно благодаря появлению новых технологий, таких как адаптированные для сетевых применений алгоритмы рассредоточения информации (IDA), алгоритмы доказательства хранения и возможности извлечения данных (PDP, POR), блокчейн-технологии. Стоит отметить, что централизованные, в том числе облачные, решения и децентрализованные хранилища приватных данных на основе P2P-сетей не исключают, а скорее взаимно дополняют друг друга. Например, если данные требуют обработки на мощных суперкомпьютерах, удобнее хранить их в централизованном (возможно, облачном) хранилище рядом (в сетевом смысле) с суперкомпьютером, на котором они будут обрабатываться. С другой стороны, если данные предполагается использовать/обрабатывать на компьютерах пользователей (например, видео- или аудиоданные для непосредственного просмотра или прослушивания)  удобнее их хранить в распределенной инфраструктуре. Другим примером ситуации, когда экономически и технически может оказаться выгодным использование децентрализованных решений является потребность в крупном хранилище на ограниченный срок осуществления какой-либо проекта, особенно в случае, когда проект объединяет многих организационно несвязанных между собой участников. Существенно важным использование децентрализованных решений является в случаях, когда характер хранимых данных предъявляет повышенные требования к приватности, отсутствию выделенных точек отказа или отсутствию выделенных компонентов, незаконное проникновение в которые позволяет осуществлять злонамеренный контроль над всей системой. Поэтому разработка полноценной технологии создания децентрализованных систем хранения приватных данных на основе P2P-сетей, передовых ИТ-методов, алгоритмов работы, протоколов и инструментария является весьма важной и актуальной.

Таким образом, теоретические исследования и практические реализации P2P-сетей различного типа являются динамично и успешной развиваемой областью ИТ-технологий с большими потенциальными возможностями разнообразных приложений. Однако, предстоит провести множество исследований и решить ряд технических проблем, чтобы распределенные хранилища приватных данных на публичных ресурсах в P2P-сетях  могли использоваться так же успешно, как это происходит в случае файлообменных систем с публичным контентом.


\label{lastpage}
\end{document}